\documentclass[sigconf]{acmart}

\AtBeginDocument{%
  \providecommand\BibTeX{{%
    \normalfont B\kern-0.5em{\scshape i\kern-0.25em b}\kern-0.8em\TeX}}}

\copyrightyear{2023}
\acmYear{2023}
\setcopyright{rightsretained}
\acmConference[KDD '23] {Proceedings of the 29th ACM SIGKDD Conference on Knowledge Discovery and Data Mining}{August 6--10, 2023}{Long Beach, CA, USA.}
\acmBooktitle{Proceedings of the 29th ACM SIGKDD Conference on Knowledge Discovery and Data Mining (KDD '23), August 6--10, 2023, Long Beach, CA, USA}
\acmPrice{}
\acmISBN{979-8-4007-0103-0/23/08}
\acmDOI{10.1145/3580305.3599473}

\usepackage{xcolor}
\usepackage{enumitem}
\usepackage{enumerate}
\usepackage{hyperref} 
\usepackage[ruled,linesnumbered]{algorithm2e}

\begin{document}

\title{PrefRec: Recommender Systems with Human Preferences for Reinforcing Long-term User Engagement}

\author{Wanqi Xue}
\authornote{The work was done during an internship at Kuaishou Technology.}
\affiliation{%
  \institution{Nanyang Technological University}
  \country{Singapore}}
\email{wanqi001@e.ntu.edu.sg}

\author{Qingpeng Cai}
\affiliation{%
  \institution{Kuaishou Technology}
  \country{Beijing, China}}
\email{caiqingpeng@kuaishou.com}

\author{Zhenghai Xue}
\affiliation{%
  \institution{Nanyang Technological University}
  \country{Singapore}}
\email{zhenghai001@e.ntu.edu.sg}

\author{Shuo Sun}
\affiliation{%
  \institution{Nanyang Technological University}
  \country{Singapore}}
\email{shuo003@e.ntu.edu.sg}

\author{Shuchang Liu}
\affiliation{%
  \institution{Kuaishou Technology}
  \country{Beijing, China}}
\email{liushuchang@kuaishou.com}

\author{Dong Zheng}
\affiliation{%
  \institution{Kuaishou Technology}
  \country{Beijing, China}}
\email{zhengdong@kuaishou.com}

\author{Peng Jiang}
\affiliation{%
  \institution{Kuaishou Technology}
  \country{Beijing, China}}
\email{jiangpeng@kuaishou.com}

\author{Kun Gai}
\affiliation{%
  \institution{Unaffiliated}
  \country{Beijing, China}}
\email{gai.kun@qq.com}

\author{Bo An}
\affiliation{%
  \institution{Nanyang Technological University}
  \country{Singapore}}
\email{boan@ntu.edu.sg}

\renewcommand{\shortauthors}{Xue et al.}

\begin{abstract}
Current advances in recommender systems have been remarkably successful in optimizing immediate engagement. However, long-term user engagement, a more desirable performance metric, remains difficult to improve. Meanwhile, recent reinforcement learning (RL) algorithms have shown their effectiveness in a variety of long-term goal optimization tasks. For this reason, RL is widely considered as a promising framework for optimizing long-term user engagement in recommendation. Though promising, the application of RL heavily relies on well-designed rewards, but designing rewards related to long-term user engagement is quite difficult. To mitigate the problem, we propose a novel paradigm, recommender systems with human preferences (or \textbf{Pre}ference-based \textbf{Rec}ommender systems), which allows RL recommender systems to learn from preferences about users’ historical behaviors rather than explicitly defined rewards. Such preferences are easily accessible through techniques such as crowdsourcing, as they do not require any expert knowledge. With PrefRec, we can fully exploit the advantages of RL in optimizing long-term goals, while avoiding complex reward engineering. PrefRec uses the preferences to automatically train a reward function in an end-to-end manner. The reward function is then used to generate learning signals to train the recommendation policy. Furthermore, we design an effective optimization method for PrefRec, which uses an additional value function, expectile regression and reward model pre-training to improve the performance. We conduct experiments on a variety of long-term user engagement optimization tasks. The results show that PrefRec significantly outperforms previous state-of-the-art methods in all the tasks.

\end{abstract}




\maketitle

\section{Introduction}
Recent recommendation systems have achieved great success in optimizing immediate engagement such as click-through rates \cite{hidasi2015session,sun2019bert4rec}. However, in real-life applications, long-term user engagement is more desirable than immediate engagement because it directly affects some important operational metrics, e.g., daily active users (DAUs) and dwell time \cite{xue2022resact}. Despite the great importance, how to effectively optimize long-term user engagement remains a significant challenge for existing recommendation algorithms. The difficulties are mainly on i) the evolving of long-term user engagement lasts for a long period; ii) factors that affect long-term user engagement are usually non-quantitative, e.g., users' satisfactory; and iii) the learning signals which are used to update our recommendation strategies are sparse, delayed, and stochastic. When trying to optimize long-term user engagement, it is very hard to relate the changes in long-term user engagement to a single recommendation \cite{wang2022surrogate}. Moreover, the sparsity in observing the evolution of long-term user engagement makes the problem even more difficult.

\begin{figure}
    \centering
    \includegraphics[width=0.6\linewidth]{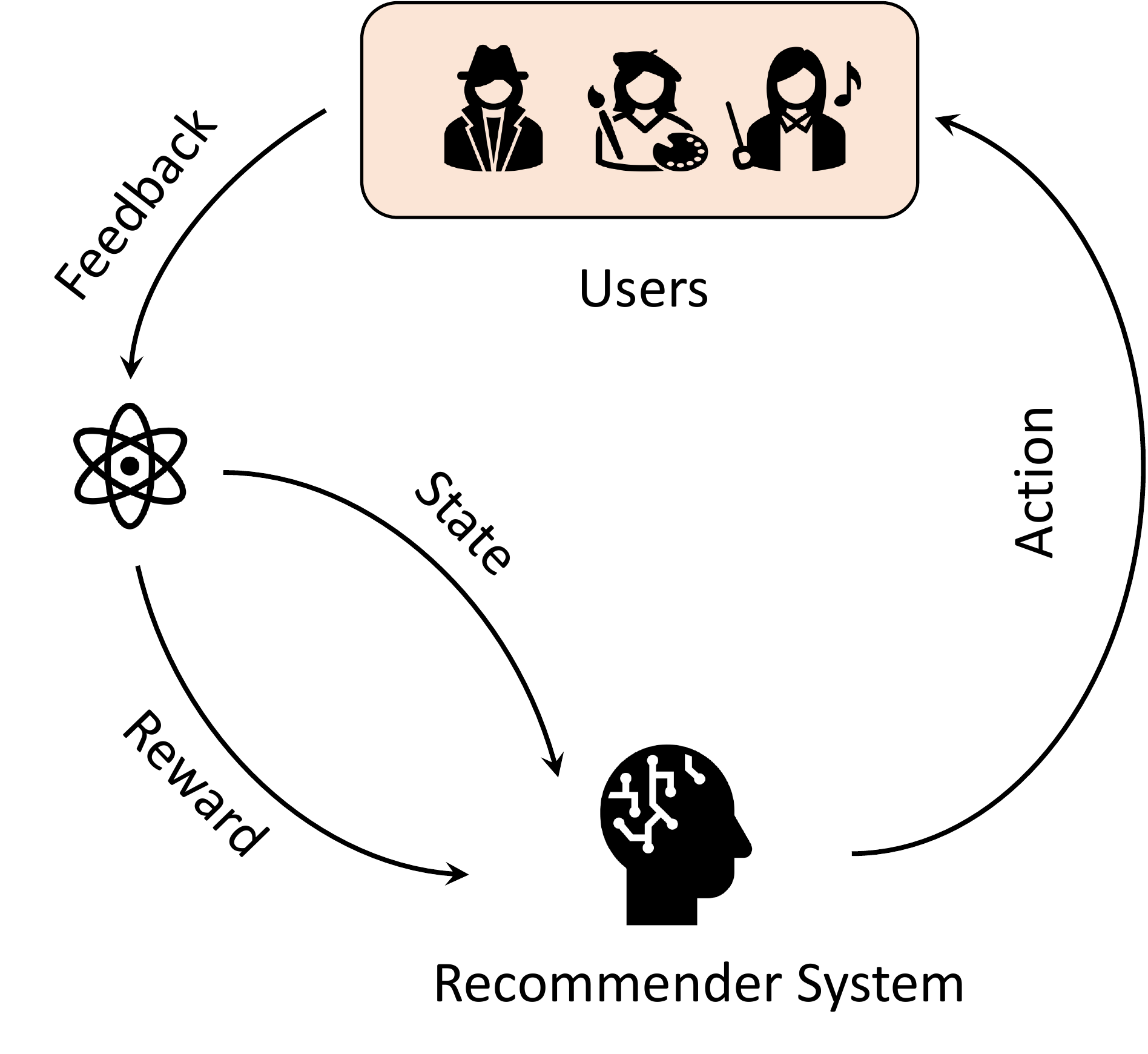}
    \caption{Reinforcement learning recommender systems.}
    \label{rl_rec}
\end{figure}

Reinforcement learning (RL) has demonstrated its effectiveness in a wide range of long-term goal optimization tasks, such as board games~\cite{alphago,alphazero}, video games~\cite{alphastar,atari}, robotics~\cite{levine2016end} and algorithmic discovery~\cite{AlphaTensor2022}.  Conventionally, when trying to solve a real-world problem with reinforcement learning, we need to first formulate it as a Markov Decision Process (MDP) and then learn an optimal policy that maximizes cumulative rewards or some other user-defined reinforcement signals in the defined MDP~\cite{sutton2018reinforcement,chen2021user}. Considering the characteristic that RL seeks to optimize cumulative rewards, it is rather suitable for optimizing long-term signals, such as user stickiness, in recommendation~\cite{zou2019reinforcement,wang2022surrogate}. As in Figure~\ref{rl_rec}, recommender systems can be modeled as an agent to interact with users which serve as the environment. Each time the agent completes a recommendation request, we can record the feedback and status changes of users to calculate a reward as well as the new state for the agent. Applying RL will lead to a recommendation policy which optimizes user engagement from a long-term perspective.

Despite that RL is an emerging and promising framework to optimize long-term engagement in recommendation, it heavily relies on a delicately designed reward function to incentivize recommender systems behave properly. However, designing an appropriate reward function is very difficult especially in large-scale complex tasks like recommendation~\cite{wang2022surrogate,christakopoulou2022reward,cqp,cqp2}. On one hand, the reward function should be aligned with our ultimate goal as much as possible. On the other hand, rewards should be sufficiently dense and instructive to provide step-by-step guidance to the agent. For immediate engagement, we can simply use metrics such as click-through rates to generate rewards~\cite{zheng2018drn,zhao2020mahrl}. Whereas for long-term engagement, the problem becomes rather difficult because attributing contributions to long-term engagement to each step is really tough. If we only assign rewards when there is a significant change in long-term engagement, learning signals could be too sparse for the agent to learn a policy. Existing RL recommender systems typically define the reward function empirically~\cite{zou2019reinforcement,xue2022resact,christakopoulou2022reward} or use short-term signals as surrogates~\cite{wang2022surrogate}, which will severely violate the aforementioned requirements of consistency and instructivity. 

{\color{black}
To mitigate the problem, we propose a new training paradigm, recommender systems with human preferences (or \textbf{Pre}ference-based \textbf{Rec}ommender systems), which allows RL recommender systems to learn from human feedback/ preferences on users’ historical behaviors rather than explicitly defined rewards. We demonstrate that RL from human preferences (or preference-based RL), a framework that has led to successful applications such as ChatGPT~\cite{ouyang2022training}, is also applicable to recommender systems. Specifically, in PrefRec, there is a (virtual) teacher giving feedback about his/her preferences on pairs of users' behaviors. 
We use the feedback (stored in a preference buffer) to automatically train a reward model which generates learning signals for recommender systems. Such preferences are easy to obtain because no expert knowledge is required, and we can use technologies such as crowdsourcing to easily gather a large number of labeled data. Furthermore, to overcome the problem that the reward model may not work well for some unseen actions, we introduce a separate value function, trained by expectile regression, to assist the training of the critic in PrefRec\footnote{PrefRec adopts the framework of actor-critic in reinforcement learning.}. Our main contributions are threefold:
\begin{itemize}
    \item We propose a new framework, recommender systems with human preferences, to optimize long-term engagement. Our method can fully exploit the advantages of reinforcement learning in optimizing long-term goals, while avoiding the complex reward engineering in reinforcement learning.
    \item We design an effective optimization method for the proposed framwork, which uses an additional value function, expectile regression and reward model pre-training to improve the performance.
    \item We collect the first reinforcement learning from human feedback (RLHF) dataset for long-term engagement optimization problem in recommendation and propose three new tasks to evaluate performance of recommender systems.
    Experimental results demonstrate that PrefRec significantly outperforms previous state-of-the-art approaches on all the tasks.
\end{itemize}}

\begin{figure*}
    \centering
    \includegraphics[width=0.8\textwidth]{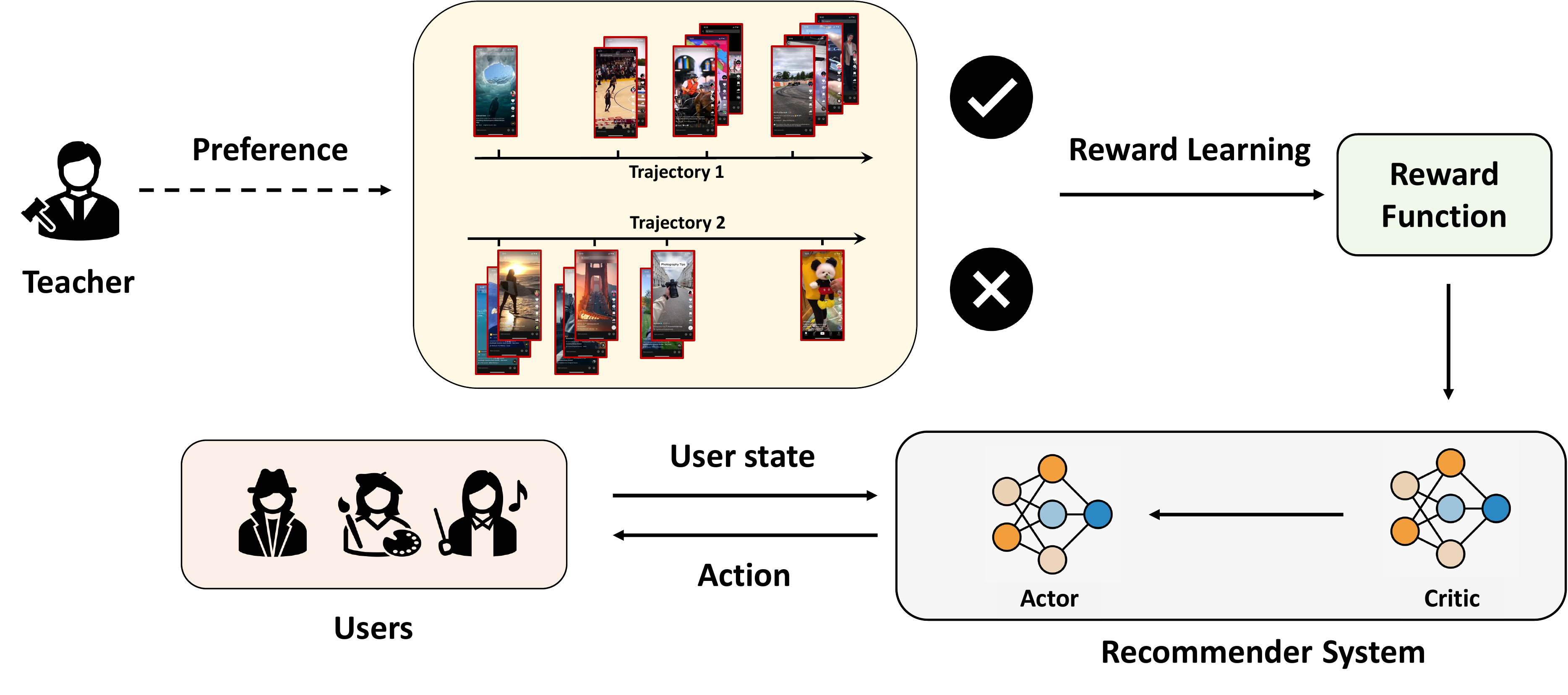}
    \caption{The framework of recommender systems with human preferences. A teacher provides feedback about his/her preferences between users' behavioral trajectories. Trajectory 1 demonstrates a trend from low-active to high-active and trajectory 2 shows an opposite tendency. Therefore, the teacher will prefer trajectory 1 to trajectory 2. With feedback of preferences, we can automatically train a reward function in an end-to-end manner. The preference-based recommender systems is then optimized by using rewards predicted by the learned reward function.}
    \label{fig:framework}
\end{figure*}

\section{Preliminaries}
\label{preliminaries}
\subsection{Long-term User Engagement in Recommendation}
\label{session-basis}
Long-term user engagement is an important metric in recommendation, and typically reflects as the stickiness of users to a product. In general, given a product, we expect users to spend more time on it and/or use it as frequently as possible. In this work, we assume that users interact with the recommender systems on a session basis: when a user accesses a product, such as an App, a session begins, and it ends when the user leaves. During each session, users can launch an arbitrary number of recommendation requests as they want. Such session-based recommender systems has been widely deployed in real-life applications such as short-form videos recommendation and news recommendation~\cite{zheng2018drn,zhan2022deconfounding,zhao2020maximizing,ijcai2019-883}. We are particularly interested in increasing
\begin{itemize}
\item[i] the number of recommended items that users consume during each visit;
\item[ii] the frequency that users visit the product.
\end{itemize}
Optimizing these two indicators is nontrivial because it is difficult to relate them to a single recommendation. For example, if a user increases its visiting frequency, we are not able to know exactly which recommendation leads to the increase. To this end, we propose to use reinforcement learning to take into account the potential impact on the future when making decisions.

\subsection{Recommendation as a Markov Decision Process (MDP)}
Applying RL to recommender systems requires defining recommendation as a Markov Decision Process (MDP). Recommender systems can be described as an agent to interact with users, who act as the environment. Formally, we formulate recommendation as a Markov Decision Process (MDP) $\langle\mathcal{S},\mathcal{A},\mathcal{P},\mathcal{R},\gamma \rangle$:
\begin{itemize}
    \item $\mathcal{S}$ is the continuous state space. $s\in\mathcal{S}$ indicates the state of a user which contains static information such as gender and age; and dynamic information, such as the rate of likes and retweets. A state is what the recommender systems relies on to make decisions.
    
    \item $\mathcal{A}$ is the continuous action space, where $a\in\mathcal{A}$ is an action which has the same dimension as the representation of recommendation items.  We determine the item to recommend by comparing the similarity of an action and item representations~\cite{zhao2020mahrl,xue2022resact}.
    
    \item $\mathcal{P}: \mathcal{S}\times\mathcal{A}\times\mathcal{S}\to\mathbb{R}$ is the transition function, where $p(s_{t+1}|s_t,a_t)$ defines the probability that the next state is $s_{t+1}$ after recommending an item $a_t$ at the current state $s_t$.
    
    \item $\mathcal{R}: \mathcal{S}\times\mathcal{A}\to\mathbb{R}$ is the reward function. $r(s_t,a_t)$ determines how much the agent will be rewarded after recommending $a_t$ at state $s_t$.
    
    \item $\gamma$ is the discount factor. $\gamma$ determines how much the agent cares about rewards in the distant future relative to those in the immediate future.
\end{itemize}
The recommendation policy $\pi(a|s): \mathcal{S}\to\mathcal{A}$ is defined as a mapping from state to action. Given a policy $\pi(a|s)$, we define a state-action value function (Q-function) $Q^{\pi}(s,a)$ which generates the expected cumulative reward (return) of taking an action $a$ at state $s$ and thereafter following $\pi$:
\begin{equation}
    Q^{\pi}(s_t,a_t)=\mathbb{E}_{(s_{t'},a_{t'})\sim \pi} \left[r(s_{t},a_{t})+\sum_{t'=t+1}^\infty \gamma^{(t'-t)}\cdot r(s_{t'},a_{t'}) \right].
\end{equation}
We seek to optimize the policy $\pi(a|s)$ so that the return obtained by the recommendation agents is maximized:
\begin{equation}
    \max_{\pi} \mathcal{J}(\pi)=\mathbb{E}_{s_t\sim d_t^{\pi}(\cdot),a_t\sim \pi(\cdot|s_t)} \left[Q^{\pi}(s_t,a_t) \right],
\end{equation}
where $d_t^{\pi}(\cdot)$ denotes the state visitation frequency at step $t$ under the policy $\pi$. By optimizing the above objective, the agent can achieve the largest cumulative return in the defined MDP. 

\begin{figure*}
    \centering
    \includegraphics[width=0.75\textwidth]{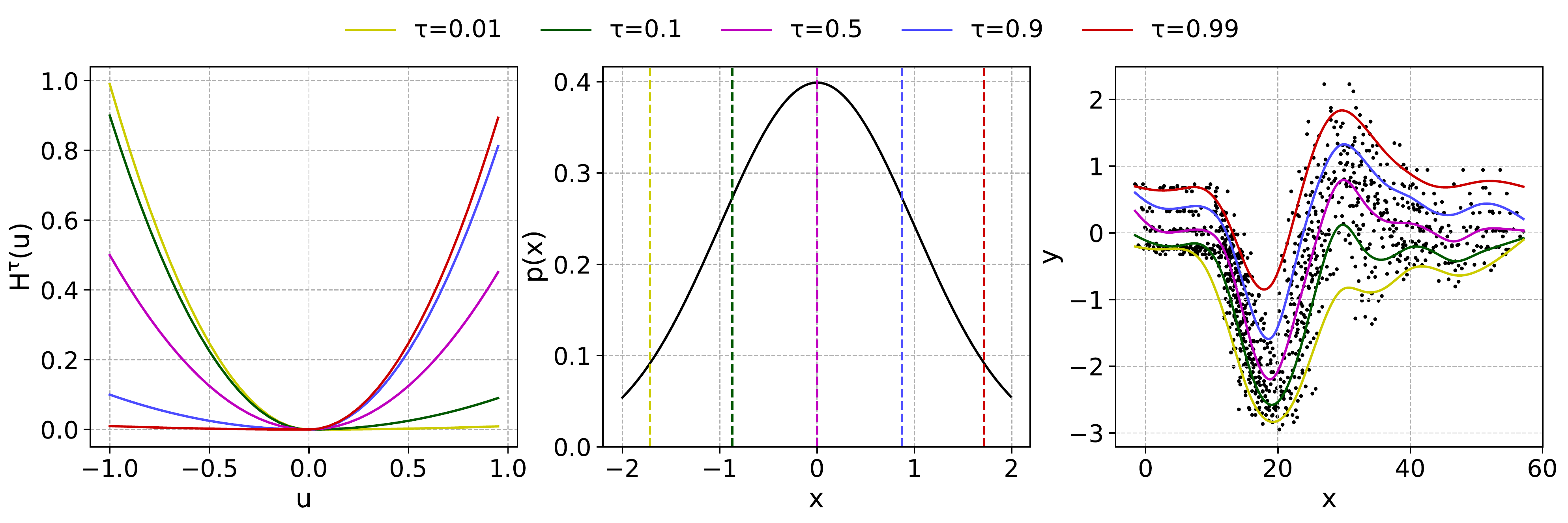}
    \caption{Left: Illustration of the expectile loss function $H^{\tau}(u)$. $\tau=0.5$ corresponds to the MSE loss, and $\tau>0.5$ upweights positives differences $u$. Center: Expectiles of a Gaussian distribution. Right: An example of expectile regression on a two-dimensional distribution.}
    \label{fig:expectile}
\end{figure*}
\section{Challenges of Designing the Reward Function}
\label{challenges}
Despite that RL is a promising approach to optimize long-term user engagement, the application of RL heavily relies on a well-designed reward function. The reward function must be able to reflect the changes in long-term engagement at each time step.
In the meantime, it should be able to provide instructive guidance to the agent for optimizing the policy.
Practically, quantifying rewards properly is very challenging because it is really difficult to relate changes in long-term engagement to a single recommendation~\cite{wang2022surrogate}. For example, when recommending a video to a user, we have no way of knowing how many videos the user will continue to watch on the platform before exiting the current session, and obviously, how this amount will be affected by the recommendation is even harder to know. For this reason, it is challenging to give a reward regarding the impact of recommended videos on the average video consumption of users. Similarly, recommending a video cannot be used to predict when the user will revisit the platform after exiting the current session. Designing rewards related to the visiting frequency of users is also difficult. As a compromise solution, one could only assign rewards at the beginning or the end of a session. However, this kind of reinforcement signals will be too sparse for the recommender systems to learn a reasonable policy, especially when the session length is large~\cite{xue2022resact}. Existing methods either design the reward function highly empirically~\cite{zou2019reinforcement,xue2022resact,christakopoulou2022reward} or use immediate engagement signals as surrogates~\cite{wang2022surrogate}, which will cause deviation between the optimization objective and the real long-term engagement. For this reason, it is urgent to propose a framework to address the difficulties in reward designing when using RL to optimize long-term user engagement.

\section{Recommender Systems with Human Preferences}
\label{methods}
To resolve the difficulties in designing the reward function, we propose a novel paradigm, \textbf{Pef}erence-based \textbf{Rec}ommender systems (PrefRec), which allows an RL recommender systems to learn from preferences on users' historical behaviors rather than explicitly defined rewards. By this way, we can overcome the problems in designing the reward function when optimizing long-term user engagement. In this section, we first introduce how to utilize preferences to generate reinforcement signals for learning a recommender systems. Then, we discuss how to optimize the performance of PrefRec by using expectile regression to better estimate the value function. Next, we propose to pre-train the reward function to stabilize the learning process. Last, we summarize the algorithm.

\subsection{Reinforcing from Preferences}
While the reward for a recommendation request is hard to obtain, preferences between two trajectories of users' behaviors are easy to determine. For example, if one trajectory shows a transition from low-active to high-active and the other shows an opposite trend or an insignificant change, we can easily indicate the preference between them. Labeling preferences does not require any expert knowledge, so we can easily use techniques such as crowdsourcing to obtain a large amount of feedback on preferences.
In PrefRec, we assume that there is a human teacher providing preferences between user’s behaviors and the recommender systems uses this feedback to perform the task. There are mainly two advantages in using preferences: i) labeling the preference between a pair of trajectories is quite simple compared to designing rewards for every step; and ii) the recommender systems is incentivized to learn the preferred behavior directly because reinforcement signals come from preferences.

We provide the framework of recommender systems with human preferences in Figure~\ref{fig:framework}. As we can find, there is a teacher providing preferences between a pair of users' behavioral trajectories\footnote{For a pair of trajectories, they may come from different users or from different periods of the same user.}. The teacher could be humans or or even a program with labeling criteria.
For trajectory 1, the user increases its visiting frequency  gradually and consumes more and more items in each session, which indicates that it is satisfied with the current recommendation policy. On the contrary, trajectory 2 shows the user is becoming less and less active, suggesting that the current recommender system should be improved to better serve the user. The teacher will obviously prefer trajectory 1 to trajectory 2. After generating such preference data, we use them to automatically train a reward function $\hat{r}(s,a;\psi)$, parameterized by $\psi$, in an end-to-end manner. The preference-based recommender systems uses the predicted reward $\hat{r}(s,a;\psi)$ rather than hand-crafted rewards to update its policy.

\begin{algorithm}
\caption{Recommender Systems with Human Preferences}
\label{algo}
\KwIn{Preference buffer $\mathcal{D}_{p}=\{(\sigma^0,\sigma^1,y)_i\}_{i=1}^{N}$,
replay buffer $\mathcal{D}_{r}=\{(s,a,s')_i\}_{i=1}^{M}$, pre-train episodes $K$}
 Initialize the reward model $r(s,a;\psi)$, the Q-function $Q(s,a;\theta)$, the V-function $V(s;\eta)$, the recommendation policy $\pi(\cdot|s;\mu)$\label{2} \\
 Set soft-update rate $\lambda$ and initialize the target Q-function $Q(s,a;\hat{\theta})$\label{target}\label{3}\\
 \For{pre-train episodes $k=1$ to $K$}{
 Sample a mini-batch of preferences $(\sigma^0,\sigma^1,y)$ from $\mathcal{D}_p$\\
 Update the reward model $r(s,a;\psi)$: $\psi \leftarrow \psi-\nabla_{\psi}\mathcal{L}_{\psi}$ (Eq.~\ref{r_loss})
 }
\While{not end training}{
Sample a mini-batch of transitions $(s,a,s')$ for $\mathcal{D}_r$\\
Label the sampled transitions by the reward model to obtain $(s,a,\hat{r},s')$\\
Update the V-function $V(s;\eta)$: $\eta \leftarrow \eta-\nabla_{\eta}\mathcal{L}^V_{\eta}$ (Eq.~\ref{vloss})\\
Update the Q-function $Q(s,a;\theta)$: $\theta \leftarrow \theta-\nabla_{\theta}\mathcal{L}^Q_{\theta}$ (Eq.~\ref{new_td}) \\
Update the recommendation policy $\pi(\cdot|s;\mu)$: $\mu \leftarrow \mu-\nabla_{\mu}\mathcal{L}^P_{\mu}$ (Eq.~\ref{piloss})\\
Soft-update the target Q-function $Q(s,a;\hat{\theta})$: $\hat{\theta}\leftarrow \lambda\theta+(1-\lambda)\hat{\theta}$\\
\If{fine-tune reward model}{
Sample a mini-batch of preferences $(\sigma^0,\sigma^1,y)$ from $\mathcal{D}_p$\\
 Update the reward model $r(s,a;\psi)$: $\psi \leftarrow \psi-\nabla_{\psi}\mathcal{L}_{\psi}$ (Eq.~\ref{r_loss})}
}
\end{algorithm}

Formally, a trajectory $\sigma$ is a sequence of observations and actions $\{(s_1,a_1),\dots,(s_T,a_T)\}$. Given a pair of trajectories $(\sigma^0, \sigma^1)$, a teacher provides a feedback indicating which trajectory is preferred, i.e., $y=(0,1)$, $(1,0)$ or $(0.5,0.5)$, where $(0,1)$ indicates trajectory $\sigma^1$ is preferred to trajectory $\sigma^0$, i.e., $\sigma^{1} \succ \sigma^{0}$; $(1,0)$ indicates $\sigma^{0} \succ \sigma^{1}$; and $(0.5,0.5)$ implies an equally preferable case. Each feedback is triple $(\sigma^0,\sigma^1,y)$ which is stored in a preference buffer $\mathcal{D}_{p}=\{((\sigma^0,\sigma^1,y))_i\}_{i=1}^{N}$. We use a deep neural network with parameters $\psi$ to predict the reward $\hat{r}(s,a;\psi)$ at a specific step $(s,a)$. By following the Bradley-Terry model~\cite{bradley1952rank,park2022surf,prefrl}, we assume that the teacher’s probability of preferring a trajectory depends exponentially on the accumulated sum of the reward over the trajectory. Then the probability that trajectory $\sigma^1$ is preferred to trajectory $\sigma^0$ can be written as a function of $\hat{r}(s,a;\psi)$:
\begin{equation}
    P_{\psi}\left[\sigma^{1} \succ \sigma^{0}\right]=\frac{\exp \left(\sum_{t} \hat{r}\left(\mathbf{s}_{t}^{1}, \mathbf{a}_{t}^{1};\psi \right)\right)}{\sum_{i \in\{0,1\}} \exp \left(\sum_{t} \hat{r}\left(\mathbf{s}_{t}^{i}, \mathbf{a}_{t}^{i};\psi \right)\right)}.
\end{equation}
Since the preference buffer $\mathcal{D}_{p}=\{((\sigma^0,\sigma^1,y))_i\}_{i=1}^{N}$ contains true labels of preferences, we can train the reward model $\hat{r}(s,a;\psi)$ through supervised learning, updating it by minimizing the cross-entropy loss:
\begin{equation}
\label{r_loss}
\mathcal{L}_{\psi}=-\underset{\left(\sigma^{0}, \sigma^{1}, y\right) \sim \mathcal{D}_p}{\mathbb{E}}\left[y(0) \log P_{\psi}\left[\sigma^{0} \succ \sigma^{1}\right]+y(1) \log P_{\psi}\left[\sigma^{1} \succ \sigma^{0}\right]\right],
\end{equation}
where $y(0)$ and $y(1)$ are the first and second element of $y$, respectively. After learning the reward model $\hat{r}(s,a;\psi)$, we can use it to generate learning signals for training the recommendation policy.

\subsection{Optimizing the Recommendation Policy}
When learning the recommendation policy, there is a replay buffer $\mathcal{D}_r$ storing training data. Unlike conventional RL recommender systems, we store $(s,a,s')$ rather than $(s,a,r,s')$ in the buffer $\mathcal{D}_r$ because we cannot obtain explicit rewards from the environment\footnote{With a slight abuse of notation, we use $s'$ to denote the next state.}. We utilize the learned reward model $\hat{r}(s,a;\psi)$ to label the reward for a tuple $(s,a)$. After labelling, an intuitive approach is to learn the Q-function by minimizing the following temporal difference (TD) error:
\begin{equation}
\label{naive_td}
\mathcal{L}^{Q}_{\theta}=\underset{\left(s,a,s'\right) \sim \mathcal{D}_r}{\mathbb{E}}\left[(\hat{r}(s,a;\psi)+\gamma Q(s',\pi(\cdot|s');\hat{\theta})-Q(s,a;\theta))^2 \right],
\end{equation}
where $\mathcal{D}_r$ is the replay buffer, $Q(s,a;\theta)$ is a parameterized Q-function, $Q(s,a;\hat{\theta})$ is a target network (e.g.,
with soft updating of parameters defined via Polyak averaging), and $\pi(\cdot|s)$ is the recommendation policy. However, in practice, we find that this method does not perform well. It may be because the recommendation policy $\pi(\cdot|s)$ will choose significantly different actions from the stored data, making the reward function and the Q-function unable to predict the corresponding values. To resolve this, we introduce a separate value function (V-function) that predicts how good or bad a state is. By doing so, we can eliminate the uncertainty that comes with the recommended policy. Instead of minimizing the TD error in Eq.~\ref{naive_td}, we turn to minimize the following loss to learn the Q-function:
\begin{equation}
\label{new_td}
\mathcal{L}^{Q}_{\theta}=\underset{\left(s,a,s'\right) \sim \mathcal{D}_r}{\mathbb{E}}\left[(\hat{r}(s,a;\psi)+\gamma V(s';\eta)-Q(s,a;\theta))^2 \right],
\end{equation}
where $V(s';\eta)$ is the V-function with parameters $\eta$. Given the replay buffer $\mathcal{D}_r$, the relationship between Q-function and V-function is
\begin{equation}
\label{v-q}
    V(s)=\mathbb{E}_a[Q(s,a)], (s,a) \in \mathcal{D}_r.
\end{equation}
Conventionally, we can optimize the parameters of the V-function by minimizing the following Mean Squared Error (MSE) loss:
\begin{equation}
    \label{vloss_0}
    \mathcal{L}^{V}_{\eta}=\underset{\left(s,a\right) \sim \mathcal{D}_r}{\mathbb{E}}\left[(Q(s,a;\hat{\theta})-V(s;\eta))^2 \right].
\end{equation}
However, such V-function corresponds to the behavior policy which collects the replay buffer $\mathcal{D}_r$. We want to achieve improvement upon the behavior policy. Inspired by expectile regression~\cite{koenker2001quantile,kostrikov2022offline}, we let the V-function to regress the $\tau$ expectile ($\tau \geq 0.5$) of $Q(s,a)$ rather than the mean statistics as in Eq.~\ref{v-q}. Then the loss for V-function becomes:
\begin{equation}
    \label{vloss}
    \mathcal{L}^{V}_{\eta}=\underset{\left(s,a\right) \sim \mathcal{D}_r}{\mathbb{E}}\left[H^{\tau}(Q(s,a;\hat{\theta})-V(s;\eta)) \right],
\end{equation}
where $H^{\tau}(u)=|\tau-\mathbb{I}(u<0)|u^2$, $\mathbb{I}(\cdot)$ is the indicator function. Specially, if $\tau=0.5$, Eq.~\ref{vloss_0} and Eq.~\ref{vloss} are identical. For $\tau > 0.5$, this asymmetric loss (Eq.~\ref{vloss}) downweights the contributions of $Q(s,a;\hat{\theta})$ smaller than $V(s;\eta)$ while giving more weights to larger values (as in Fig.~\ref{fig:expectile}, left). Fig.~\ref{fig:expectile} (right) illustrates expectile regression on a two-dimensional distribution: increasing $\tau$ leads to more data points below the regression curve. Back to the learning of the V-function, the purpose is to let $V(s;\eta)$ to regress an above-average value. The Q-function is jointly trained with the V-function, while it also serves as the critic to guide the update of the recommendation policy, i.e., the actor. The recommendation policy is optimized by minimizing the loss:
\begin{equation}
    \label{piloss}
    \mathcal{L}^{P}_{\mu}=- \underset{\left(s,a\right) \sim \mathcal{D}_r}{\mathbb{E}}\left[Q(s, \pi(\cdot|s;\mu);\theta)\right],
\end{equation}
where $\pi(\cdot,s;\mu)$ is the recommendation policy with parameters $\mu$.

\subsection{Pre-training the Reward Function}
The reward function is used to automatically generate reinforcement signals to train the recommendation policy. However, A reward model that is not well-trained will cause the collapse of the recommendation policy. Thus, to stabilize the training, we propose to pre-train the reward function before starting the updating of the recommendation policy. Specifically, we first prepare a preference buffer that stores a pool of preference feedback between interaction histories. Then we initialize a deep neural network with the structure of multi-layer perceptron to learn from the preference buffer. The neural network is updated by minimizing the loss in Eq.~\ref{r_loss}. After training for several episodes, we start the training of the recommendation policy. At this phase, the reward model is updated simultaneously with the recommender systems. By doing so, the reward model can handle potential shift in human preferences and can therefore generate more accurate learning signals.

\subsection{Overall Algorithm}
We provide the overall algorithm of PrefRec in Algorithm~\ref{algo}. A preference buffer $\mathcal{D}_p$ and a replay buffer $\mathcal{D}_r$ is required for training PrefRec. From lines 1 to 2, we do the initialization for the deep neural networks and set hyper-parameters such as soft-update rate. From lines 3 to 5, we pre-train the reward model to confirm that it is able to provide reasonable learning signals when updating the recommendation policy. Lines 7 to 18 describe the training process of the recommendation policy. We first sample a mini-batch of transitions from the replay buffer $\mathcal{D}_r$. Then we lable the samples data with the reward model. Following that, we update the parameters of the V-function, the Q-function and the recommendation policy accordingly. Finally, if fine-tuning the reward model, we will train the reward model to keep it up-to-date.
{\color{black}
\subsection{Discussions}

PrefRec differs from both inverse reinforcement learning (IRL)~\cite{10.5555/645529.657801} and model-based RL~\cite{Kaiser2020Model}. The key differences between PrefRec and IRL include that IRL requires costly expert demonstrations while PrefRec only requires simple label-like feedback. The reward function in PrefRec is learned by aligning with human preferences, whereas in IRL it is inferred from expert demonstrations. On the other hand, model-based RL relies on environmental rewards and aims to simplify learning by approximating the reward and transition functions. In contrast, PrefRec does not require environmental rewards and is designed to be learned without them.
}
\section{Experimental Results}
\label{exp}
We conduct extensive experiments to evaluate our algorithm. In particular, we will answer the following research questions (RQs):
\begin{itemize}
    \item \textbf{RQ1}: Can the framework of PrefRec lead to improvement in long-term user engagement?
    \item \textbf{RQ2}: Whether PrefRec is able to outperform existing state-of-the-art methods?
    \item \textbf{RQ3}: Whether the learned reward signals is able to reflect the true underlying rewards?
    \item \textbf{RQ4}: How do the components in PrefRec contribute to the performance? 
\end{itemize}


\begin{figure}
    \centering
    \includegraphics[width=0.45\textwidth]{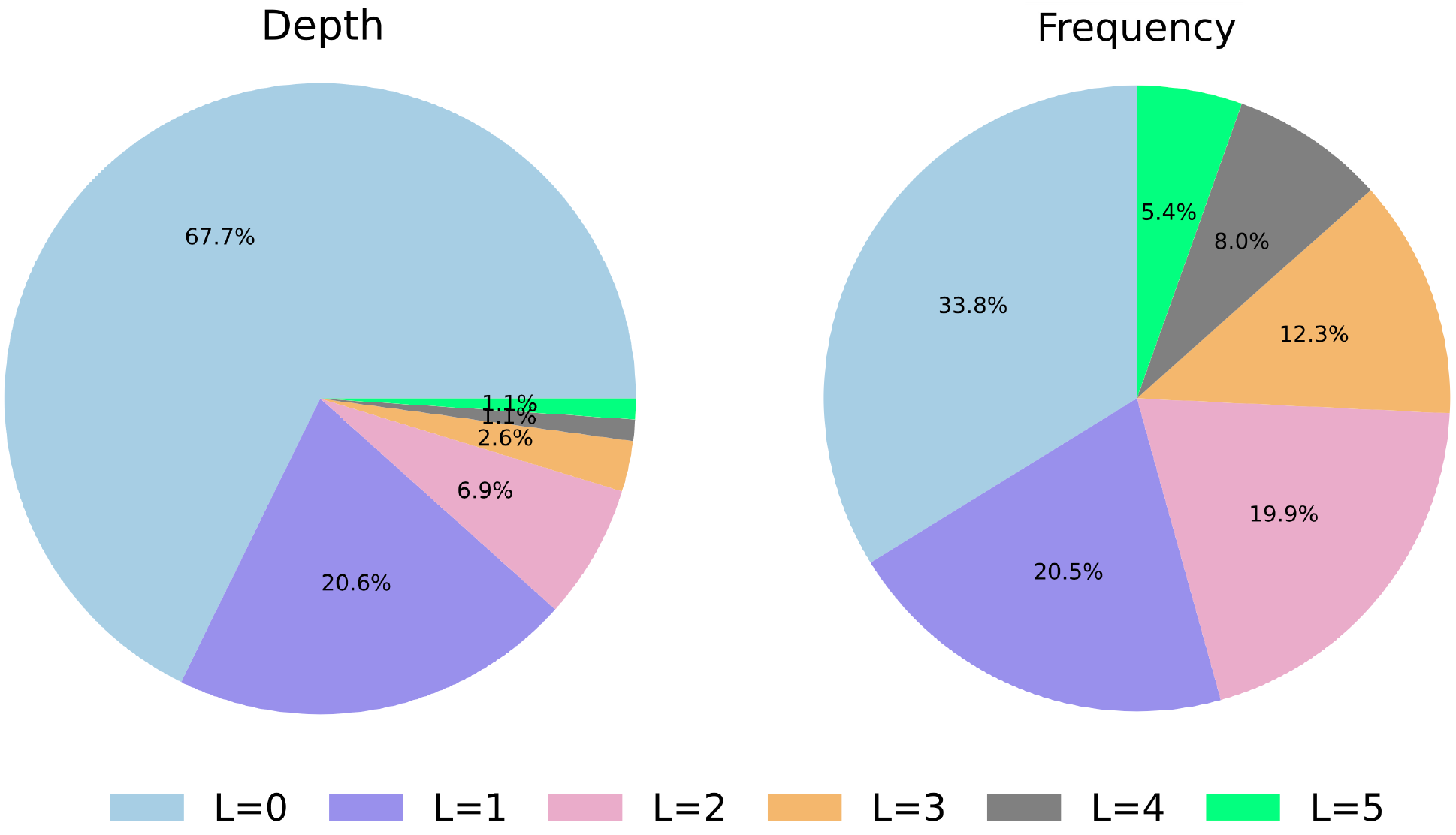}
    \caption{The proportion of the levels in session depth and visiting frequency.}
    \label{fig:l_stat}
\end{figure}

\begin{figure*}
    \centering
    \includegraphics[width=0.8\textwidth]{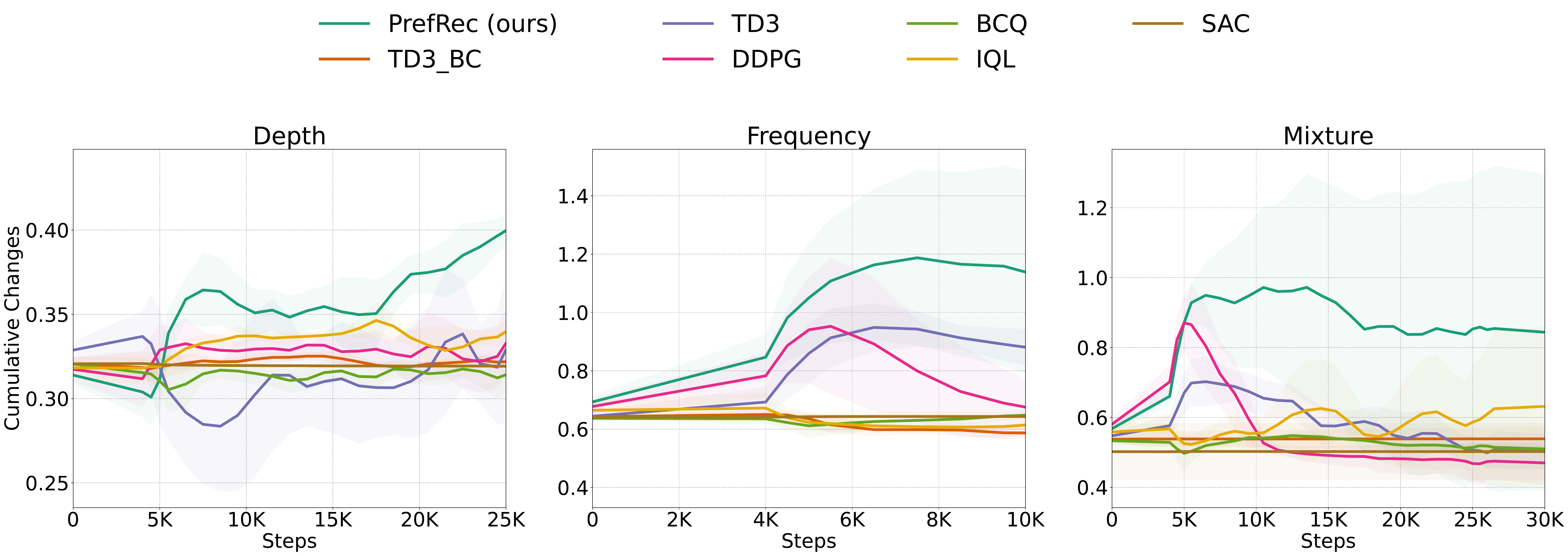}
    \caption{Learning curves of RL recommender systems under the framework of PrefRec, averaged over 5 runs. }
    \label{fig:learning_curves}
\end{figure*}

\subsection{Preparing the dataset}
Since PrefRec is a new framework in recommendation, there is no available dataset for evaluation. To prepare the dataset, we track complete interaction histories of around $100,000$ users from a leading short-form videos platform for months, during which over 25 millions recommendation services are provided\footnote{Data samples and codes are at \href{https://www.dropbox.com/sh/hgsqg5fabnvmp26/AABF-2dvarI_bdyygYEt5aw7a?dl=0}{\nolinkurl{https://www.dropbox.com/sh/hgsqg5fabnvmp26/AABF-2dvarI_bdyygYEt5aw7a?dl=0}}.}. For each user, we record its interaction history in a session-request form: the interaction history consists of several sessions with each session containing a number of recommendation requests (see Sec.~\ref{session-basis}). The recommender system provides service at each recommendation request where we record the state of a user and the action of the recommender system. Each state is a 245-dimensional vector containing the user's state information, such as gender, age and historical like rate. We applied scaling, normalization and clipping to each dimension of states in order to stabilize the input of models. An action is a 8-dimensional vector which is the representation of recommended item at a request. We record the timestamp of each time a user starts and exits a session. With the timestamp, we can calculate the duration that a user revisits the platform and thus can infer the visiting frequency.
We measure changes in long-term user engagement at the end of each session for each user. Specifically, we first calculate the average session depth $\delta^u_{avg}$ and the average revisiting time $\epsilon^u_{avg}$ for each user $u$ during the time span. For each session, $\delta^u_i$ denotes the number of requests in the $i$-th session of the user $u$. If $\delta^u_i$ is larger than the average session depth $\delta^u_{avg}$, we consider that there is an improvement in session depth. Similarly, we can calculate the revisiting time $\epsilon^u_{i}$ for session $i$ and if it is less than the average revisiting time $\epsilon^u_{avg}$, we consider that there is an improvement in visiting frequency. We quantify the changes in session depth and visiting frequency into six levels (level 0 to level 5; the more higher level, the more positive change) where the levels are determined by the following equations, respectively:
\begin{equation}
\begin{aligned}
\label{level}
    \mathbf{L}^u_{i}(depth)=\left(\lfloor \frac{\delta^u_i}{\delta^u_{avg}} \rfloor\right).clip(0,5) \\
     \mathbf{L}^u_{i}(frequency)=\left(\lfloor \frac{\epsilon^u_{avg}}{\epsilon^u_{i}} \rfloor\right).clip(0,5)
\end{aligned}
\end{equation}
We provide the proportion of the calculated levels of all sessions in Fig.~\ref{fig:l_stat}. For session depth, most of sessions stay in level $0$ and only very few of them are located in level $4$ and $5$. The visiting frequency demonstrates a more even distribution.

After processing the data, we can prepare the two buffers, i.e., the preference buffer $\mathcal{D}_p$ and the replay buffer $\mathcal{D}_r$, which are used in PrefRec. For the preference buffer $\mathcal{D}_p$, we uniformly sample $20,000$ pairs of users who have launched for more than $200$ recommendation requests in the platform. We set the length of trajectory $\sigma$ as $100$ and randomly sample a segment of trajectories with this length from the interaction histories of the selected users. To generate the preferences, we write a scripted teacher who provides its feedback by comparing cumulative levels of changes on the trajectory segments. For the replay buffer $\mathcal{D}_r$, we randomly sample $80\%$ of users as the training set and split their interaction histories into transitions $(s,a,s')$to fill up the replay buffer $\mathcal{D}_r$.

\subsection{Baselines}
We compare PrefRec with a variety of baselines, including reinforcement learning methods for off-policy continues control (DDPG, TD3, SAC), offline reinforcement learning algorithms (TD3\_BC, BCQ, IQL), and imitation learning:
\begin{itemize}
    \item \textbf{DDPG}~\citep{lillicrap2015continuous}: An off-policy reinforcement learning algorithm which concurrently learns a Q-function and a deterministic policy. The update of the policy is guided by the Q-function.
    \item \textbf{TD3}~\citep{fujimoto2018addressing}: A reinforcement learning algorithm which is designed upon DDPG. It applies techniques such as clipped double-Q learning, delayed policy updates, and target policy smoothing.
    \item \textbf{SAC}~\citep{haarnoja2018soft}:  An off-policy reinforcement learning algorithm trained to maximize a trade-off between expected return and entropy, a measure of randomness in the policy. It also incorporates tricks such as the clipped double-Q learning.
    \item \textbf{TD3\_BC}~\citep{fujimoto2021minimalist}: An offline reinforcement learning algorithm designed based on TD3. It applies a behavior cloning (BC) term to regularize the updating of the policy.
    \item \textbf{BCQ}~\citep{fujimoto2019off}: An offline algorithm which restricts the action space in order to force the agent towards behaving similarly to the behavior policy. 
    \item \textbf{IQL}~\citep{kostrikov2022offline}: An offline reinforcement learning method which uses expectile  regression to estimate the value of the best action in a state.
    \item \textbf{IL}: Imitation learning treats the behaviors in the replay buffer as expert knowledge and learns a mapping from observations to actions by using expert knowledge as supervisory signals.
\end{itemize}
{\color{black}
Since our work focuses on addressing complex reward engineering when reinforcing long-term engagement and how to convey human intentions to RL-based recommender systems, we mainly make comparisons with classical RL algorithms. Works like FeedRec~\cite{zou2019reinforcement} emphasizes on designing DNN architecture and is orthogonal to PrefRec which focuses on the policy optimization process.}

\subsection{Evaluation Metric}
Among the $10,0000$ users, we sample $80\%$ of them as the training set and the remaining $20\%$ users constitute the test set. For the test users, we store their complete interaction histories separately, in the session-request format. We adopt Normalised Capped Importance Sampling (NCIS)~\cite{swaminathan2015self}, a widely used standard offline evaluation method~\cite{gilotte2018offline,farajtabar2018more}, to evaluate the performance. Formally, the score of a policy $\pi$ is calculated by
\begin{equation}
\begin{aligned}
\tilde{J}^{NCIS}(\pi)=\frac{1}{|\mathcal{U}|}\sum_{u \in \mathcal{U}}\left[ \frac{\sum_{i=0}^{|\mathcal{T}_u|}\tilde{\rho}_i(\pi,\mathcal{T}_u)\mathbf{L}^u_i}{\sum_{i=0}^{|\mathcal{T}_u|}\tilde{\rho}_i(\pi,\mathcal{T}_u)}\right],
\end{aligned}
\end{equation}
where $\mathcal{U}$ is the set of test users, $\mathcal{T}_u$ is the set of sessions of the user $u$, $\tilde{\rho}_i(\pi,\mathcal{T}_u)$ is the probability that the policy $\pi$ follows the request trajectory of the $i$-th session in $\mathcal{T}_u$, and $\mathbf{L}^u_i$ is the level of change for the $i$-the session (as defined in Eq.~\ref{level}). Intuitively, $\tilde{J}^{NCIS}(\pi)$ awards a policy with a high score if the policy has large probability to follow a good trajectory.
{\color{black}
\subsection{Implementation Details}
\label{exp_detail}
To ensure fairness in comparison across all methods and experiments, a consistent network architecture is utilized. This architecture consists of a 3-layer Multi-Layer Perceptron (MLP) with 256 neurons in each hidden layer. The hyper-parameters for the PrefRec method are listed in Table~\ref{hyper}. All methods were implemented using the PyTorch framework.

\begin{table}[ht]
    \centering
    \caption{Hyper-parameters of PrefRec.}
    \scalebox{1}{
    \begin{tabular}{l|c}
    \toprule[1pt]
        \textbf{Hyper-parameter} &  \textbf{Value} \\ \midrule[0.5pt]
        Optimizer & Adam~\citep{kingma2014adam} \\
        Actor Learning Rate & $5\times 10^{-6}$ \\
        Critic Learning Rate & $5\times 10^{-5}$ \\
        State Dimensions & 245 \\
        Action Dimensions & 8 \\
        Transitions Batch Size & 4096 \\
        Preferences Batch Size & 256 \\
        Normalized Observations & Ture \\
        Gradient Clipping & False\\
        Fine-Tuning & True \\
        Discount Factor & 0.9 \\
        Expectile Rate & 0.7 \\
        Soft-update Rate & $0.999$ \\ 
        Segment Length & 100 \\
        Preference Buffer Size & $2 \times10^{4} $ \\
        Replay Buffer Size & $3 \times10^{6} $ \\
        Number of Pre-train Epoch & 3 \\
        Number of Train Epoch & 5 \\ \bottomrule[1pt]
    \end{tabular}
    }
    \vspace{-0.25cm}
    \label{hyper}
\end{table}
}

\subsection{Overall Results}
We conduct experiments to verify if the framework of PrefRec can improve long-term user engagement in terms of i) session depth; ii) visiting frequency; and iii) a mixture of the both. We randomly sample $500$ users from the test set and use them to plot the learning curves of our method and the baselines. As in Fig.~\ref{fig:learning_curves}, PrefRec achieves a significant and consistent increase in cumulative long-term engagement changes in all the tasks, although it does not receive any explicit reinforcement signal. Similar phenomenon can also be observed in some of those generic reinforcement learning algorithms, such as DDPG. They demonstrates growth in specific tasks, though not that stable. The learning curves indicate that the learned reward function is able to provide reasonable reinforcement signals. and the framework of PrefRec provides an effective training paradigm to achieve reward-free recommendation policy learning. Next, we save the models with the best performance in training and test their performance on the whole test set. As in Table~\ref{tbl:overall_performance}, those generic reinforcement learning algorithms poorly without explicit rewards. PrefRec is able to outperform all the baselines by a wide margin in all the three tasks, showing the effectiveness of the proposed optimization methods. 
\begin{table}
\centering
\caption{Overall performance comparisons on various long-term user engagement optimization tasks. The ``$\pm$\small'' indicates $95\%$
    confidence intervals.}
\scalebox{0.95}{
\begin{tabular}{lccc}
\toprule[1pt] 
 & Depth  & Frequency  & Mixture   \\
  \midrule[0.5pt]
 DDPG~\citep{lillicrap2015continuous} & 0.3211 $\pm$\small0.0060 & 1.1408 $\pm$\small0.0163 & 1.0642 $\pm$\small0.0119\\
 TD3~\citep{fujimoto2018addressing} & 0.3495 $\pm$\small 0.0038 & 0.9714 $\pm$\small 0.0141 & 0.6423 $\pm$\small 0.0097\\
 SAC~\citep{haarnoja2018soft} & 0.3190 $\pm$\small 0.0031 & 0.6416 $\pm$\small 0.0058 & 0.5348 $\pm$\small 0.0047\\ \midrule[0.5pt]
 TD3\_BC~\citep{fujimoto2021minimalist} & 0.3246 $\pm$\small 0.0032 & 0.6781 $\pm$\small 0.0060 & 0.5326 $\pm$\small 0.0047\\
 BCQ~\citep{fujimoto2019off} & 0.3142 $\pm$\small 0.0033 & 0.6580 $\pm$\small 0.0060 & 0.5528 $\pm$\small 0.0052\\
 IQL~\citep{kostrikov2022offline} & 0.3311 $\pm$\small 0.0054 & 1.0354 $\pm$\small 0.0141 & 0.8230 $\pm$\small 0.0099\\ \midrule[0.5pt]
 IL & 0.3202 $\pm$\small 0.0031 & 0.6406 $\pm$\small 0.0058 & 0.5348 $\pm$\small 0.0047\\  \midrule[0.5pt]
 PrefRec (ours) &\textbf{0.4229 $\pm$\small0.0077} &\textbf{1.7706 $\pm$\small0.0206} &\textbf{1.3788 $\pm$\small0.0133}\\ 
 $\%$ Improv.& 21.00$\%$& 55.20$\%$ & 29.56$\%$\\ \bottomrule[1pt]
\end{tabular}
}
\label{tbl:overall_performance}
\end{table}
Despite utilizing only a single dataset, optimizing session depth and visiting frequency are distinct challenges. The results of the experiments demonstrate the generalization ability of PrefRec as it consistently delivers improvements across both tasks. The optimization of session depth and visiting frequency are not necessarily interdependent; the algorithm that produces a deep session may not result in high visiting frequency, and similarly, high visiting frequency does not guarantee a deep session (as seen in Fig.~\ref{fig:learning_curves}).

\begin{figure}[ht]
    \centering
    \includegraphics[width=0.48\textwidth]{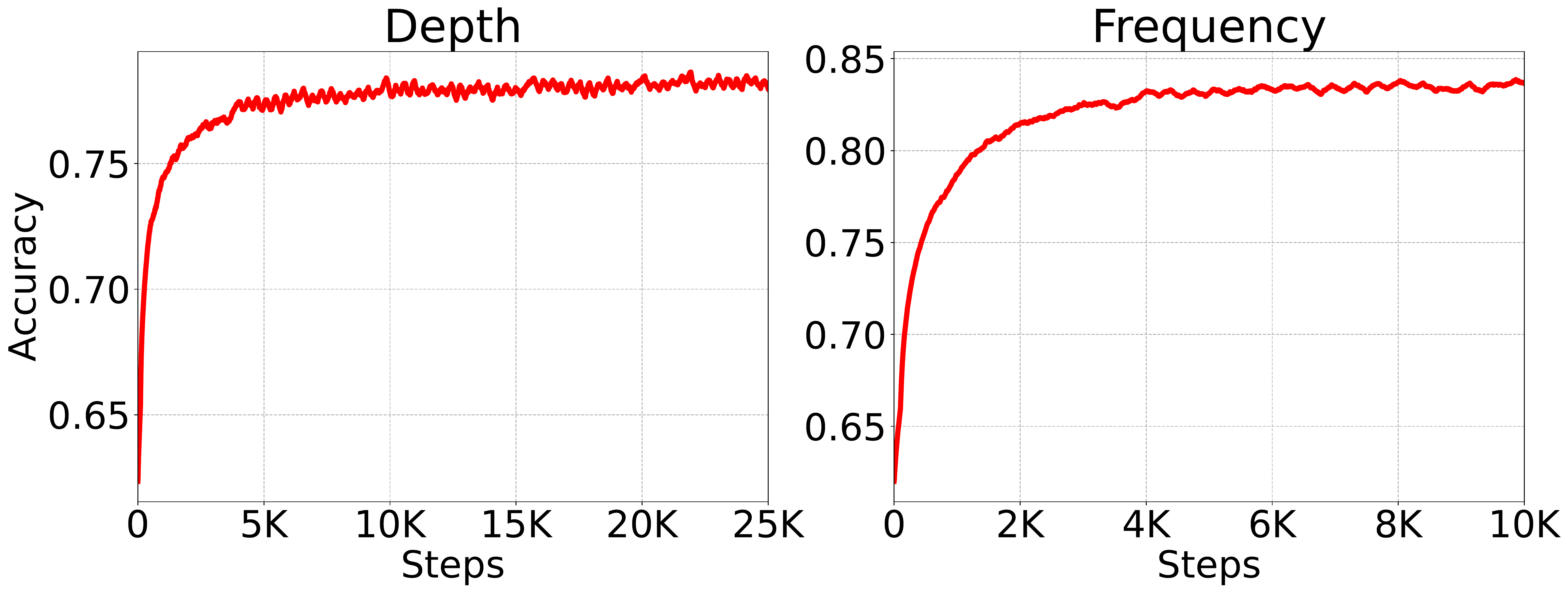}
    \caption{Prediction accuracy of the reward function.}
    \label{fig:r_accuracy}
\end{figure}
\begin{figure}[ht]
    \centering
    \includegraphics[width=0.48\textwidth]{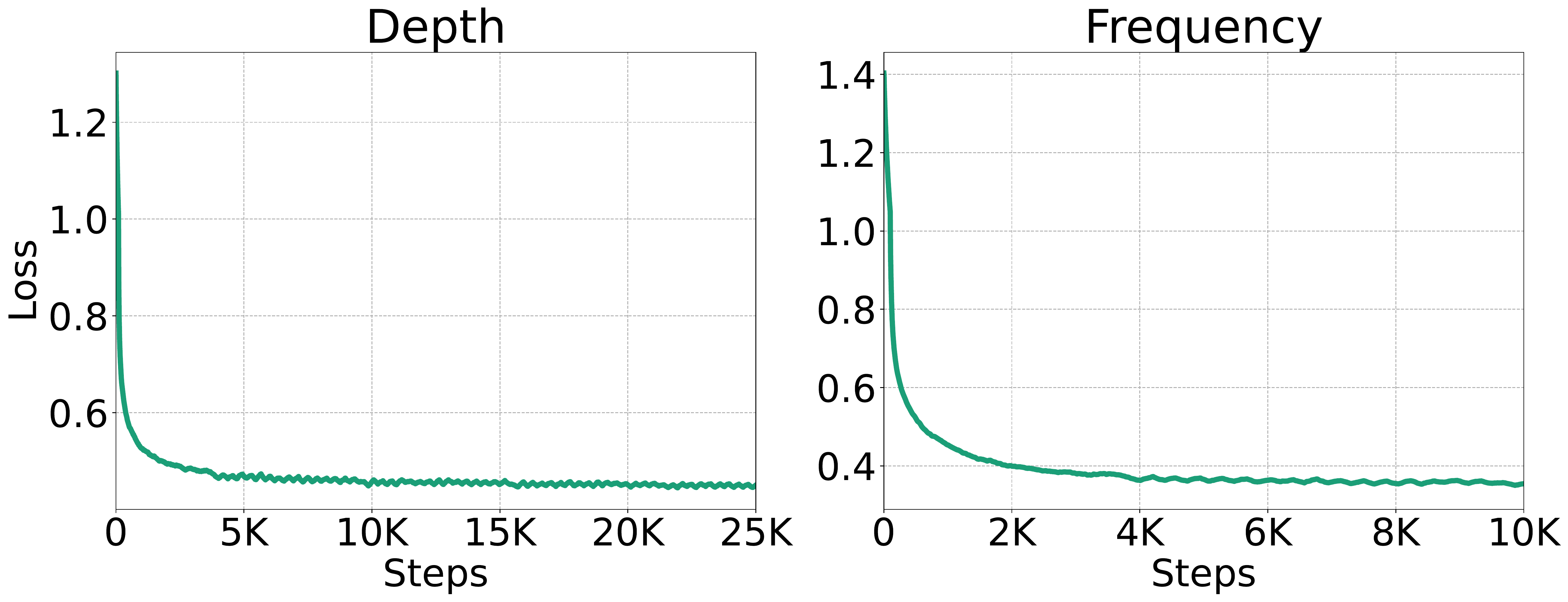}
    \caption{Learning loss of the reward function.}
    \vspace{-0.25cm}
    \label{fig:r_loss}
\end{figure}

{\color{black}
\subsection{Training Process of the Reward Function}
\label{r_plot}
To ensure the validity of the learning signals generated by the reward function, it must accurately predict the preferences that are utilized in the training phase. To evaluate the performance of the reward function, we have plotted its prediction accuracy in Fig.~\ref{fig:r_accuracy}. The results show a noticeable improvement in accuracy as the training process advances. Additionally, we also plot the learning loss of the reward function in Fig.~\ref{fig:r_loss}. The results indicate a substantial decrease in the loss, reducing it to a much lower level compared to its initial value.
The improvement in prediction accuracy and the reduction in loss demonstrate that the reward function is becoming increasingly effective at accurately predicting the preferences used in the training phase. This is crucial for ensuring that the learning signals generated by the reward function are reliable and accurate, enabling the model to learn from the preferences effectively.
}

\begin{figure}[ht]
    \centering
    \includegraphics[width=0.48\textwidth]{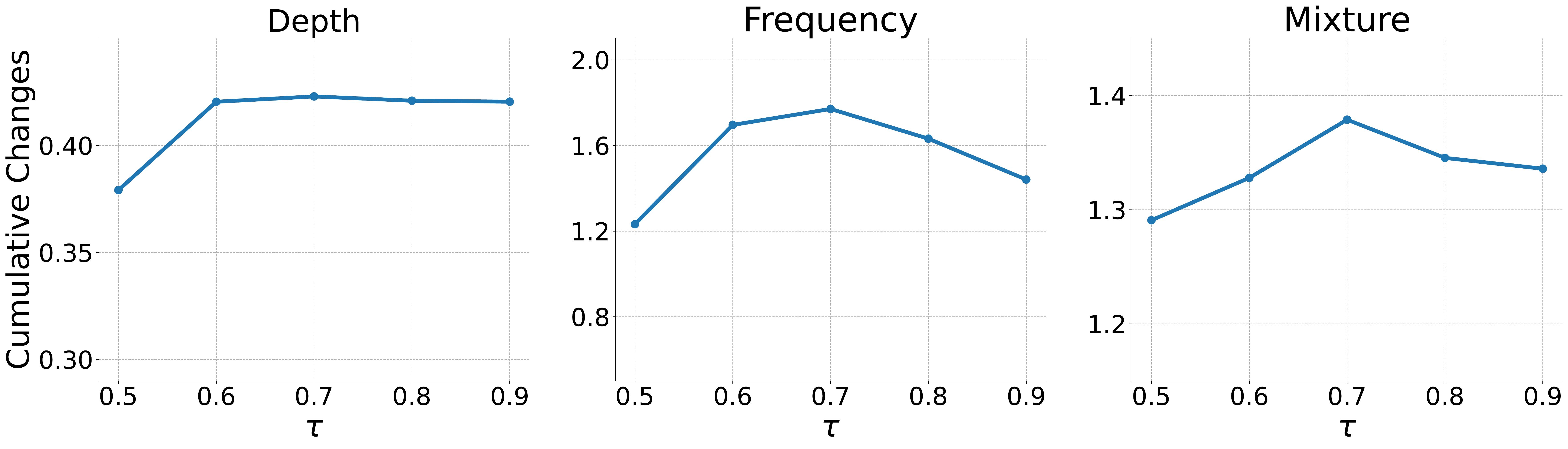}
    \caption{Ablations on the expectile regression factor $\tau$.}
    \label{fig:tau_ab}
\end{figure}

\begin{figure}[ht]
    \centering
    \includegraphics[width=0.48\textwidth]{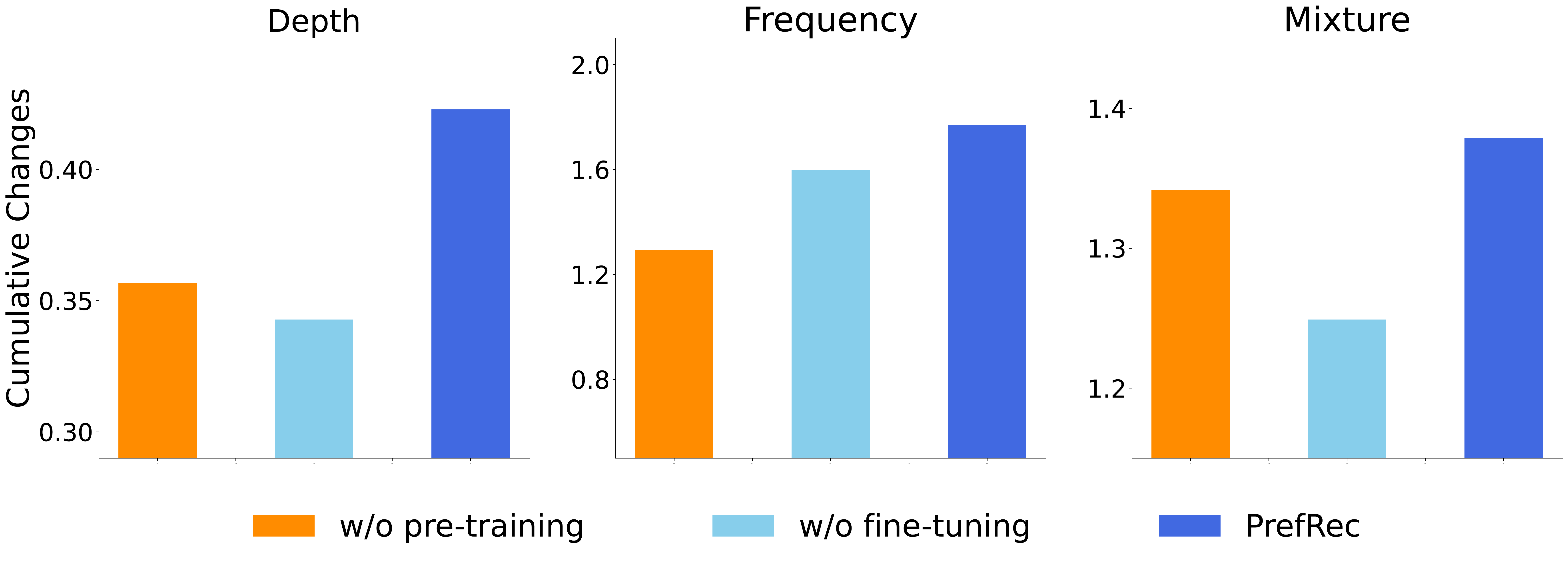}
    \caption{Ablations on reward function pre-training and reward function fine-tuning.}
    \vspace{-0.25cm}
    \label{fig:pretrian_ab}
\end{figure}

\subsection{Ablations}
To understand how the components in PrefRec affect the performance, we perform ablation studies on the expectile regression factor and the pre-training/fine-tuning of the reward function. As can be found in Table~\ref{fig:tau_ab}, when $\tau=0.5$ where the expectile regression becomes identical to the regression to mean, there is an obvious drop in performance. The phenomenon indicates that the expectile regression with a proper $\tau$ contributes to the improvement in performance. What's more, if compared with the results of DDPG in Table~\ref{tbl:overall_performance}, we can find that introducing a separate V-function also benefits the performance since DDPG can be considered an algorithm whose $\tau=0.5$ and without a V-function. Next, we study whether the pre-training and fine-tuning of the reward function is useful. As in Table~\ref{fig:pretrian_ab}, if we directly update the recommendation policy without pre-training the reward function, the performance decreases. Similarly, training the recommendation policy with a frozen reward function also degrades the performance. The results suggests that the pre-training and fine-tuning of the reward function are important to the performance.

\section{Related Work}
\label{relatedwork}
\subsection{Optimizing Long-term User Engagement in Recommendation}
Long-term user engagement has recently attracted increasing attention due to its close connection to user stickiness. One of the major difficulties in promoting long-term user engagement is the lack of consistent and informative reward signals. Reward signals can be sparse and change over time, making it difficult to accurately predict and respond to user needs.  Zhang et al.~\citep{zhang2021counterfactual} augments the sparse rewards by re-weighting the original feedbacks with counterfactual importance sampling. Wu et al.~\citep{wu2017returning} consider user click as immediate reward and user's return time as a hint on long-term engagement. It also assumes a stationary distribution of candidates for recommendation. In contrast, Zhao et al.~\citep{zhao2021rabbit} empirically identify the problem of non-stationary users' taste distribution and propose distribution-aware recommendation methods. Other researches focus on maximizing the diversity of recommendations as an indirect approach to optimize long-term engagement. For example, Adomavicius et al.~\citep{adomavicius2011maximizing} propose a graph-based approach
to maximize diversities based on maximum flow. Ashton et al.~\citep{ashton2020algorithmic} propose that high recommendation diversities is related to long-term user engagement. Apart from diversities, Cai et al.~\citep{cqp2} conduct large-scale empirical studies and propose several surrogate criteria for optimizing long-term user engagements, including high-quality consumption, repeated consumption, etc. In this paper we avoid directly learning from the sparse and non-stationary signals inferred from the environment. Instead, we propose to learn a parameterized reward function to optimize the long-term user engagement.
\subsection{Reinforcement Learning for Recommender Systems}
Reinforcement learning allows an autonomous agent to interact with environment and optimizes long-term goals from experiences, which is particularly suitable for tasks in recommender systems~\citep{zou2019reinforcement,mazoure2021improving,shani2005An,bai2019model,chen2018stabilizing,zhang2022multi}. Shani et al.~\citep{shani2005An} first proposed to employ Markov Decision Process~(MDP) to model the recommendation behavior. The subsequent researches focus on standard RL problems, e.g., exploration and exploitation trade-off in the context of recommendation systems~\citep{chen2021values,tang2017exploration}, together with model-based or simulator-based approaches to improve sample efficiency~\citep{shi2019virtual,bai2019model}. Recently, there has been works on designing proper reward functions for efficient training of recommender systems. For example, Zou et al.~\citep{zou2019reinforcement} use a Q-network with hierachical LSTM to model both the instant and long-term reward. Ji et al.~\citep{ji2021reinforcement} model the recommendation process as a Partially Observable MDP and estimate the lifetime values of the recommended items. Zheng et al.~\citep{zheng2018drn} explicitly model the future returns by introducing the user activeness score. Ie et al.~\citep{ie2019slateq} decompose the Q-function for tractable and more efficient optimization. There are also researches focusing on behavior diversity~\citep{wang2022surrogate,zhou2010solving} as a surrogate for the reward function. Instead of relying on the aforementioned handcrafted reward signals, in this paper we propose to automatically train a reward function based on preferences between users' behavioral trajectories, which avoids the difficulties in reward engineering. 
\section{Conclusions}
\label{conclusion}
In this paper, we propose PrefRec, a novel paradigm of recommender systems, to improve long-term user engagement. PrefRec allows RL recommender systems to learn from preferences between users' historical behaviors rather than explicitly defined rewards. By this way, we can fully exploit the advantages of RL in optimizing
long-term goals, while avoiding complex reward engineering. PrefRec uses the preferences to automatically learn a reward function. Then the reward function is applied to generate reinforcement signals for training the recommendation policy. We design an effective optimization method for PrefRec, which utilizes an additional value function, expectile regress and reward function pre-training to enhance the performance. Experiments demonstrate that PrefRec significantly and consistently outperforms the current state-of-the-art on various of long-term user engagement optimization tasks.
\begin{acks}
This research is supported by the National Research Foundation, Singapore under its Industry Alignment Fund – Pre-positioning (IAF-PP) Funding Initiative. Any opinions, findings and conclusions or recommendations expressed in this material are those of the author(s) and do not reflect the views of National Research Foundation, Singapore.
\end{acks}

\bibliographystyle{ACM-Reference-Format}
\bibliography{sample-base}


\end{document}